# Is Platinum a Proton Blocking Catalyst?

*Aparna Saksena*[*a], *Yujun Zhao*[a], *J. Manoj Prabhakar*[a], *Dierk Raabe*[a], *Baptiste Gault*[a,b], *Yug Joshi*[*a]

[a] Max Planck Institute for Sustainable Materials, Max-Planck-Straβe 1, Düsseldorf 40237, Germany

[b] Department of Materials, Royal School of Mines, Imperial College London, Prince Consort Road, London, SW7 2BP, UK

*corresponding author: a.saksena@mpi-susmat.de, y.joshi@mpi-susmat.de

## Abstract

Platinum, to date, is the most widely applied electrocatalyst for hydrogen evolution reaction (HER) in acidic media. It is assumed to be a proton-blocking catalyst with only surface-limited adsorption of the reaction intermediates. Here, we critically evaluate the bulk interaction of Pt with hydrogen (H), and its heavier isotope deuterium (H/D), by monitoring *operando* mass change of the Pt electrode during galvanostatic heavy/water splitting by employing an electrochemical quartz crystal microbalance. Unexpectedly, we observe an irreversible temporal mass gain and a change in the reaction's overpotential, arising from diffusion of H/D into Pt, confirmed by atom probe tomography and thermal desorption spectroscopy. Sub-surface concentration of at least ca. 15 at. % of D in Pt was observed, diffusing down to a depth of more than 10 nm. Analytical description quantified the diffusion coefficient of D in Pt to be $(3.2\pm0.05)\times10^{-18}$ $cm^2 \cdot s^{-1}$. These findings challenge the existing credence of Pt-proton interaction being limited to the surface, prompting the expansion of the catalyst design strategies to account for property-modifying bulk diffusion of H/D in the Pt matrix.

Introduction

Hydrogen is poised to become the favored carbon-free alternative energy vector[1]. It can be sustainably produced by using renewable energy for water electrolysis, where the hydrogen evolution reaction (HER) occurs at the cathode[2]. Catalysts reduce the activation energy barrier[3] of the reaction and therefore, their design is critical to increase hydrogen-production efficiency and rate. The Sabatier principle suggests that the performance of a catalyst scales with the binding energy of the reactants that need to adsorb on the surface and the products to desorb[4]. The change in the produced current density as a function of the binding energy of $H/H_2$ on a material's surface forms the well-known "volcano plot"[5].

To date, platinum is the most widely used and remains the most active catalyst for HER[6] in acidic media. Pt is very close to the optimum balance between adsorption and desorption, i.e., the interaction of H is neither too weak nor too strong, enabling it to reach the highest conversion of $H^+$ from the solution to gaseous $H_2$. However, the limited abundance and expensive synthesis of Pt drives up costs, limits the scaling up of hydrogen-based technologies, and motivates the efficiency of its usage. There is hence a critical need to elucidate the mechanisms underlying both its exceptional catalytic efficiency and its degradation, so as to define design strategies that maximize HER performance and durability. In this context, we make a surprising discovery that the actual surface reaction energetics that was assumed to stand behind this efficiency of Pt as a catalyst, work differently than assumed so far.

At positive potentials with respect to the standard hydrogen electrode[7], hydrogen was reported to adsorb on the Pt surface, often termed as hydrogen underpotential deposition. This is typically observed as a characteristic reduction in current at underpotential[8]. Also, for HER, Pt is reported to be a proton blocking catalyst[9], with no diffusion of H into Pt being expected[5] and interactions assumed to be limited to the very surface. Previous permeation tests conducted on Pt showed an extremely low solubility of 1 hydrogen atom in $10^{11}$ atoms of Pt[10] at 25°C, which further strengthens the belief that H cannot diffuse into the bulk of Pt and can only be adsorbed at the surface, which had never been experimentally verified. Recently, a "hydride-like" reaction intermediate, forming on the top 4 monolayers of Pt was reported.[11] However, the formation of this hydride can be attributed to the alkaline species in the electrolyte, causing cathodic corrosion as reported in prior studies[12].

Here, we evidence a mass gain of the electrode during hydrogen/deuterium evolution reaction (HER/DER) by operando electrochemical quartz crystal microbalance (eQCM). eQCM precisely quantifies time-resolved loading and unloading of mass of an electrode *operando*, as showcased in batteries[13-15], supercapacitors[16], and gas sensing[17]. Following DER, atom probe tomography (APT) analysis of the Pt thin film electrode confirms that deuterium not just adsorbs on the surface but diffuses into the bulk Pt during the reaction. This unsuspected ingress of H/D deep below the surface during the reaction will inherently affect the local or surface electrochemical potential[18]. Bulk diffusion will be more prominent with a maximized surface area to volume ratio. Finally, H could affect the durability of electrodes as hydrogen embrittles most metals. This new critical piece of the complex puzzle should motivate a rethink of the electrocatalyst's design principles.

## Results

A 3-electrode cell is used for monitoring the reaction. The Pt working electrode (WE) is placed in the electrolyte, in an H type cell where a glassy carbon electrode is used as counter electrode (CE) and is separated by a Nafion N-117 membrane, schematic representation of which is shown in Fig. 1A. A reversible hydrogen electrode (RHE) is used as a reference electrode (RE). The yellow colored holder in Fig. 1A is the WE, it connects the Pt-coated quartz crystal to the potentiostat as well as to the crystal oscillator, such that the eQCM can record the change in the resonance frequency that is directly related to the change of mass as per the Sauerbrey relation[19] (see methods section).

The cross-sectional transmission electron microscopy (TEM) image of the quartz crystal is shown in Fig. 1B. It shows a bright field image of the stack comprising the quartz ($SiO_2$) crystal substrate, a 70 nm Au thin film, and the electrochemically active 110 nm Pt thin-film, separated by Ti adhesion layers. Columnar grain can be observed for both Au and Pt films. Fig. 1C shows the selected area electron diffraction pattern obtained from the white circular region highlighted in Fig. 1B. The diffraction spots correspond to a face-centered cubic phase with a lattice parameter of 0.39 ± 0.01 nm, consistent with pure Pt[20]. Only the top Pt-film of the electrode layer is exposed to the electrolyte.

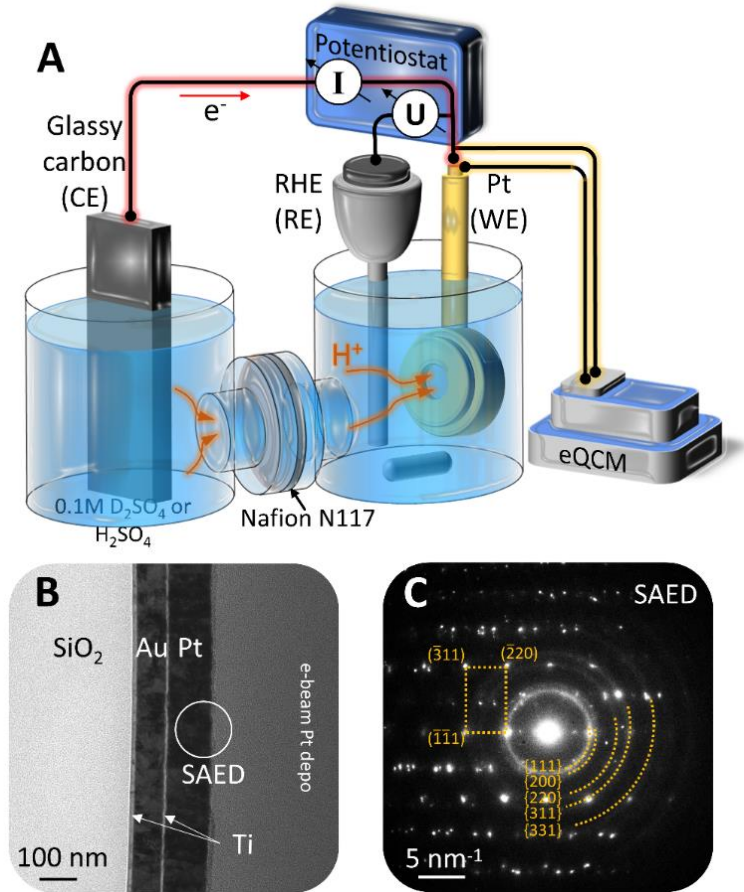

Fig. 1: A) A schematic of the 3-electrode cell used for HER reaction. Platinum-coated quartz crystals are placed in contact with 0.1 M $H_2SO_4$ and the oscillation frequency is monitored during the water and heavy water splitting reaction. **TEM of the multilayered electrode**: B) shows the bright field image of the Pt film deposited on Au coated quartz crystal where the thin film stack of Ti as an adhesion layer, Au, Ti and Pt is clearly visible. C) shows the selected area diffraction pattern of the area highlighted in B). The diffraction spots marked with the rectangular frame correspond to the Pt grains orientated along $[112]_{fcc}$ zone axis. The diffraction rings match the crystal structure of Pt, originating from the e-beam deposition

For HER and DER experiments in 0.1 M $H_2SO_4$ and 0.1 M $D_2SO_4$, a constant current density of −2.4 mA·cm$^{-2}$ is applied for 12 h, and subsequently, we record the open circuit voltage (OCV) for another 2 h (see Methods). These samples are hereafter referred as Pt$_{HER}$ and Pt$_{DER}$ respectively, and the potential evolution is plotted in grey and red in Fig. 2A for both. The voltage U steadily decreases with time, thereby increasing the overpotential (calculated beyond 0 V vs RHE) for both HER and DER. The overpotential for DER is higher than that for HER, which can be ascribed to a slower kinetics as previously reported[21, 22]. During the OCV stage, the potential immediately rises, because of the release of the transient surface adsorbed species as also has been observed in the literature[23, 24].

The resonant frequency during the electrochemical reaction is plotted as a function of time in Fig. 2B, evidencing an overall decrease for both $Pt_{HER}$ and $Pt_{DER}$, corresponding to a parabolic mass gain as observed in Fig. 2C for $Pt_{HER}$ over the 12 h of reaction. Due to more bubbling during DER, the mass change of $Pt_{DER}$ is skewed in the initial 400 s, however, subsequently, a parabolic increase in mass is also observed. With the mass of D being twice that of H[25], the rate of mass increase during DER is higher than in HER. To have a better comparison on the kinetics in HER and DER, the mass per unit area is converted to concentration per unit area in Fig. 2D. At the same current density, the bulk diffusion of deuterium appears slower than hydrogen, which is to be expected as the charge radius of D is more than double of H[26]. After HER/DER, when the electrode is held at OCV, the reversible mass change is recovered from the surface adsorbed species as the potential rises. However, Fig. 2C and 2D reveal that an irreversible mass gain of 2.9 µg·cm$^{-2}$ and 1.8 µg·cm$^{-2}$ remains in the bulk of the electrode after 12 h of HER and DER respectively.

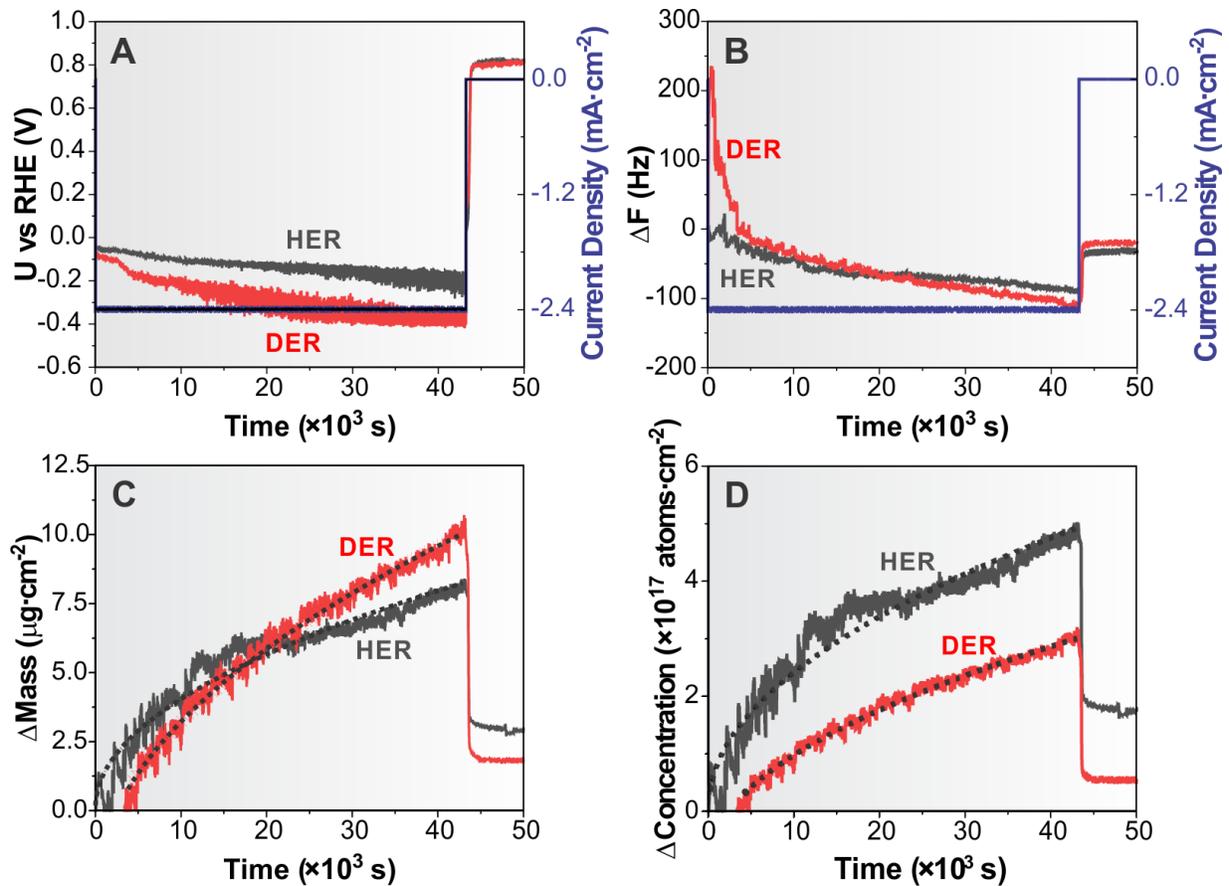

Fig. 2: Mass and potential change during HER and DER: A) shows the potential evolution durign the galvanostatic evolution of H/D. B) shows the correponsing resonance frequency change in the quartz crystal during the galvanostatic HER and DER by grey and red lines, respectively. C) the mass change in the bulk of the electrode duing HER/DER calculated from B. D) the absolute change of concentration of D and H atoms going inside the bulk quartz calculated from C.

APT is performed to spatially resolve the distribution of this irreversible D within $Pt_{DER}$. The mass-to-charge spectrum in Fig 3A reveals peaks between 1 and 4 Da pertaining to $H^+$, $H_2^+/D^+$, $H_3^+/DH^+$ and $D_2^+$ along with $Pt^{2+}$ and $Pt^+$. H is the most abundant element, and it can be incorporated into the material during specimen preparation[27] and is also present as residual gas even in the ultra-high vacuum chamber, making its detection inevitable during APT analysis[28, 29]. Depending on the magnitude of the electrostatic field during the APT measurement, H can be detected at 1 Da as $H^+$ as well as at 2 and 3 Da as $H_2^+$ and $H_3^+$, respectively, which overlaps with $D^+$ and $DH^+$, making even D quantification challenging[29, 30]. Yet, with a natural isotopic abundance of D of only ca. 0.015%[31], the amplitude of the peak at 4 Da irrefutably corresponds to $D_2^+$ and therefore proves diffusion of D in Pt during DER.

Fig. 3B displays a side view of the 3D reconstructed point cloud, in which the Pt is represented as lilac dots and the $D_2^+$ ions as red spheres, along the film thickness of the electrode. To deconvolute the contribution of H and D to the peaks at 2 and 3 Da, we performed reference measurements on Pt specimens extracted from similar thin films (see Fig. S1). The evaporation field was systematically varied as monitored by the ratio of $Pt^{1+}$ versus $Pt^{2+}$.[32] This allowed us to estimate the expected contributions from molecular $H_2^+$ and $H_3^+$ (see Fig. S2), including, for the evaporation field conditions used in the analyses of $Pt_{DER}$. These can then be subtracted to precisely quantify D (see details in supplementary information). Fig. 3C plots the one-dimensional concentration profile into the bulk of the electrode extracted, along the arrow in Fig. 3B, corrected from these contributions. The D concentration forms a plateau within the top 2.5 to 3 nm of the measurement volume at a concentration of ca. 15 at.%, followed by a drop to background levels within the first 10 nm. D losses can be expected through desorption from the surface of the $Pt_{DER}$ during sample transfer and specimen preparation, and the specimen's apex may or may not correspond to the $Pt_{DER}$ surface itself. Therefore, we most likely underestimate the D concentration in the electrode's sub-surface region, which can, in principle, be extrapolated to an error function-like profile (black dotted line in Fig 3C) provided there is no outwards diffusion during sample preparation.

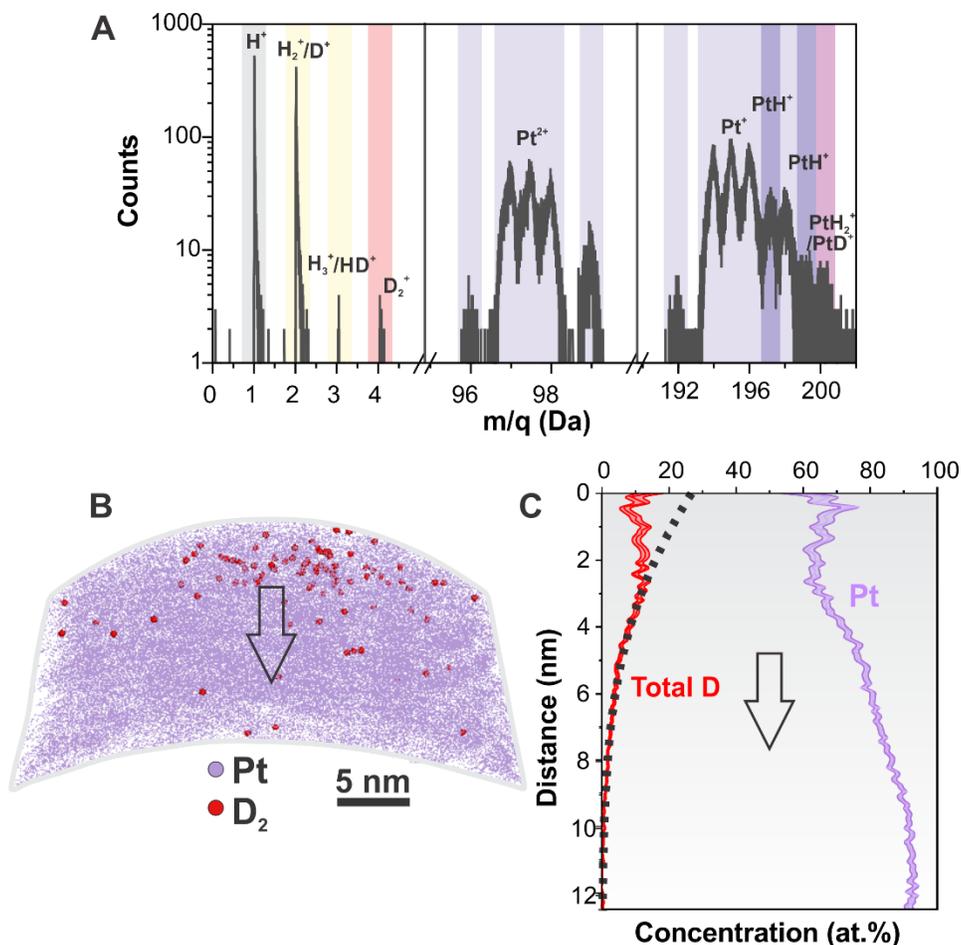

Fig. 3: **Atom Probe Tomography of Pt$_{DER}$**: A) shows the mass spectrum obtained during the measurement clearly indicating the presence of D$_2^+$ along with Pt$^{2+}$ and Pt$^+$ ; B) shows the APT reconstruction of Pt thin film after heavy water splitting where the red dots correspond to the distribution of D$^{2+}$ along the thin film thickness. C shows the one dimensional concentration profile along the direction highlighted in B. The total D is estimated after corrections made from reference measurements shown in supplementary information.

For Pt$_{HER}$, since quantification with APT is challenging, a qualitative assessment is done using thermal desorption spectroscopy (TDS), shown in Fig. S3 in the supplementary information of pristine Pt and Pt$_{HER}$ to corroborate our findings of mass gain observed during operando eQCM. The results indicate a significant increase in hydrogen desorption after the HER reaction compared to TDS performed on pristine Pt. All the measurements herein prove that there is a much higher bulk solubility of deuterium and hydrogen in Pt than conventionally assumed.

## Discussion

In HER, the Volmer step has been considered as the rate-determining step[33], but it depends on the localized hydrogen concentration. In a blocking electrode theory, only the surface modification[34] and edge effects[35] were considered. However, our experimental result points to a sub-surface concentration of at least ca. 15 at. % of deuterium going into the Pt bulk, reaching down to a depth of more than 10 nm. Since galvanostatic DER was performed, which resulted in $D_2$ evolution, it is reasonable to assume that a constant concentration of D is achieved at the Pt surface. Such a concentration-gradient driven diffusion can be described by an error-function provided a semi-infinite diffusion condition is met. This condition seems to be the case here as the concentration goes almost to zero after 10 nm. The error function fit is shown by the black dashed line in Fig. 3C. From the fitting we can quantify the diffusion coefficient of D in Pt to be $(3.2\pm0.05)\times10^{-18}$ $cm^2 \cdot s^{-1}$ and a surface composition of 30 at. % of D in Pt (see details in supplementary information).

These findings challenge the established view of Pt as a proton blocking catalyst hypothesized from the reported extremely low solubility of H in Pt[10]. From the composition, the prior reported, Pt-hydride-like phase formed during HER in alkaline condition[11] was not observed here. As a consequence, H is not confined to the surface or subsurface of Pt, it diffuses deeper (up to 10 nm ) into the Pt bulk, likely forming an interstitial solid-solution of H/D and Pt. Hence, hydrogen will introduce strain and changes in atomic neighborhoods that trigger changes in the electronic structure, which in turn governs the electrochemical activity[36]. These must now be integrated into the theoretical considerations to explain the factors limiting hydrogen production[37] for instance, and in the design principles for future Pt-based electrocatalysts.

## Methods

**Operando Electrochemistry**

The platinum electrodes were received from Advanced Wave Sensors, S.L. The operando electrochemistry was performed with two types of electrolytes i.e. 0.1 M $H_2SO_4$ (97–99% purity, Sigma Aldrich) in deionized water and 0.1 M $D_2SO_4$ (97–99% purity, Sigma Aldrich) in $D_2O$ (99% purity, Sigma Aldrich). The operando measurement of mass change was performed by employing an electrochemical quartz crystal microbalance (AW-QSD-300: BluQCM QSD-300 from Bio-Logic Science Instruments GmbH), which was combined with a SP 200 potentiostat also from Bio-Logic Science Instruments GmbH. The galvanostatic measurements were performed at a current density of -2.4 mA·cm$^{-2}$. A glassy carbon electrode (SynLectro™ by Sigma Aldrich) was used as a counter electrode, and a reversible hydrogen electrode (RHE) (Gasktel GmbH, supplied from Bio-Logic Science Instruments GmbH) was used as a reference electrode. The Quartz crystal was connected to the oscillator and to the potentiostat using a special holder (AW-PEQ11 sold by Bio-Logic Science Instruments GmbH). An H-type cell (Ossila BV) was used to perform the HER and DER reaction, as shown in the schematic in Fig. 1A. The two compartments of the H-cell are separated using a Nafion N-117 membrane (Redoxme AB), enabling only proton exchange and not graphitic or carbon species[38].

**Mass calculation using electrochemical quartz crystal microbalance (eQCM)**

The mass change per unit area from QCM or eQCM can be determined using the well-known Sauerbrey equation[19] given in Eq. 1.

$$\Delta m_Q = - \frac{\sqrt{\rho \cdot u}}{2n \cdot f^2} \cdot \Delta f = -\Delta f \cdot C_f^{-1} \qquad 1$$

where $\Delta f$ is the change of the resonance frequency, $f$ is the fundamental resonance frequency (5 MHz), ρ (2.65 g·cm$^{-3}$) is the density of the quartz, $u$ represents the shear modulus of the quartz and $n$ is the overtone order (here, it is 1). $C_f$ is a calibrated sensitivity factor 10.9188 Hz cm$^2$·µg$^{-1}$·of the used QCM sensor.

The calibrated sensitivity factor was determined by electroplating copper on eQCM crystal using 10mM copper (II) sulphate solution in 1M H$_2$SO$_4$ as electrolyte with Pt as counter electrode and reversible hydrogen electrode as the reference electrode. A cyclic voltammetry (CV) measurement within the voltage window of -0.05 V to 0.5 V vs RHE at a scan rate of 50mV·s$^{-1}$ was used for the plating and removal. The current from the CV measurement was integrated to determine the charge. The corresponding frequency change was determined using the eQCM oscillator and is plotted in Fig. S4.

Correlating this with Faraday's law of electrolysis, we can determine the calibration factor $C_f$ (in Hz-cm$^2$·µg$^{-1}$) as per Eq. 2:

$$C_f = \frac{\Delta f/Q \cdot F \cdot A \cdot n}{M_{Cu} \cdot 10^6} \qquad 2$$

Where $\Delta f/Q$ is the slope from Fig. S4, $F$ is the Faraday Constant, $A$ is the mass sensitive area in the Quartz Crystal (0.1963 cm$^2$), $n$ is valence of the charged species (in this case +2) and $M_{Cu}$ is the molecular mass of electrodeposited species or Copper in this case (63.54 g·mol$^{-1}$).

**Atom Probe Tomography**

For analyzing the local composition of the Pt$_{DER}$, APT was performed on a local electrode atom probe (LEAP) 5000XR. Laser-assisted field evaporation at a laser frequency of 125 kHz and laser energy of 70 to 150 pJ was used to detect, on average, 5 ions per 1000 pulses at 60 K. The APT specimens were prepared using a FEI dual-beam focused ion beam (FIB) Helios 600i using the standard lift out process outlined in reference[39]. The sample transfer from the electrochemical setup to the FIB and then subsequently to the APT is performed in ambient conditions, where the APT specimen preparation in the FIB is performed at ambient temperature without intentional heating/ cooling.

**Transmission Electron Microscopy**

Transmission electron microscopy (TEM) images, and selected area electron diffraction (SAED) patterns were recorded using a JEOL JEM-2100PLUS. For this, electron-transparent foils were prepared using the FIB lift-out procedure[40] with Thermo Fisher Helios 5 CX Dual-Beam Ga FIB/SEM.

**Thermal Desorption Spectroscopy**

Thermal Desorption Spectroscopy (TDS) experiments were carried out using a Hiden TPD workstation integrated with a mass spectrometer to measure the desorbed hydrogen from the pristine Pt sample and the sample that underwent HER. The spectra were measured at a constant heating rate of 8°C min$^{-1}$ from room temperature to 600 °C

# Acknowledgement

The authors gratefully acknowledge Uwe Tezins, Andreas Sturm, Christian Bross, and Volker Kree for their support to the FIB, APT, and TEM facilities at MPI-SusMat. AS and BG acknowledge the financial support from Deutsche Forschungsgemeinschaft (DFG) under the collaborative research centre CRC 1625 under project B4. Also, AS and BG acknowledge discussions with Bingxin Li and Mira Todorova. YZ acknowledges funding by the European Union through the EIC grant no. 101184347, Heat2Battery. YJ would like to acknowledge critical discussions with Dr. Yongqiang Kang and acknowledge funding by the European Union's research and innovation programme Horizon Europe under the grant agreement No. 101192848, and also the Max-Planck Society.## References

1. Kovač, A.; Paranos, M.; Marciuš, D., Hydrogen in energy transition: A review. *International Journal of Hydrogen Energy* **2021,** *46* (16), 10016-10035.
2. Dotan, H.; Landman, A.; Sheehan, S. W.; Malviya, K. D.; Shter, G. E.; Grave, D. A.; Arzi, Z.; Yehudai, N.; Halabi, M.; Gal, N.; Hadari, N.; Cohen, C.; Rothschild, A.; Grader, G. S., Decoupled hydrogen and oxygen evolution by a two-step electrochemical-chemical cycle for efficient overall water splitting. *Nature Energy* **2019,** *4* (9), 786-795.
3. Wang, S.; Lu, A.; Zhong, C. J., Hydrogen production from water electrolysis: role of catalysts. *Nano Converg* **2021,** *8* (1), 4.
4. Laursen, A. B.; Varela, A. S.; Dionigi, F.; Fanchiu, H.; Miller, C.; Trinhammer, O. L.; Rossmeisl, J.; Dahl, S., Electrochemical Hydrogen Evolution: Sabatier's Principle and the Volcano Plot. *Journal of Chemical Education* **2012,** *89* (12), 1595-1599.
5. Parsons, R., The Rate of Electrolytic Hydrogen Evolution and the Heat of Adsorption of Hydrogen. *Transactions of the Faraday Society* **1958,** *54* (7), 1053-1063.
6. Hansen, J. N.; Prats, H.; Toudahl, K. K.; Morch Secher, N.; Chan, K.; Kibsgaard, J.; Chorkendorff, I., Is There Anything Better than Pt for HER? *ACS Energy Lett* **2021,** *6* (4), 1175-1180.
7. Jerkiewicz, G., Standard and Reversible Hydrogen Electrodes: Theory, Design, Operation, and Applications. *Acs Catalysis* **2020,** *10* (15), 8409-8417.
8. Osawa, M.; Tsushima, M.; Mogami, H.; Samjeské, G.; Yamakata, A., Structure of water at the electrified platinum-water interface:: A study by surface-enhanced infrared absorption spectroscopy. *Journal of Physical Chemistry C* **2008,** *112* (11), 4248-4256.
9. Spencer, M. A.; Holzapfel, N. P.; You, K. E.; Mpourmpakis, G.; Augustyn, V., Participation of electrochemically inserted protons in the hydrogen evolution reaction on tungsten oxides. *Chem Sci* **2024,** *15* (14), 5385-5402.
10. Ebisuzaki, Y.; Kass, W. J.; O'Keeffe, M., Solubility and Diffusion of Hydrogen and Deuterium in Platinum. *The Journal of Chemical Physics* **1968,** *49* (8), 3329-3332.
11. Hersbach, T. J. P.; Garcia-Esparza, A. T.; Hanselman, S.; Paredes Mellone, O. A.; Hoogenboom, T.; McCrum, I. T.; Anastasiadou, D.; Feaster, J. T.; Jaramillo, T. F.; Vinson, J.; Kroll, T.; Garcia, A. C.; Krtil, P.; Sokaras, D.; Koper, M. T. M., Platinum hydride formation during cathodic corrosion in aqueous solutions. *Nat Mater* **2025,** *24* (4), 574-580.
12. Hersbach, T. J. P.; McCrum, I. T.; Anastasiadou, D.; Wever, R.; Calle-Vallejo, F.; Koper, M. T. M., Alkali Metal Cation Effects in Structuring Pt, Rh, and Au Surfaces through Cathodic Corrosion. *ACS Appl Mater Interfaces* **2018,** *10* (45), 39363-39379.
13. Wang, K.; Joshi, Y.; Chen, H.; Schmitz, G., In-situ analysis of solid-electrolyte interphase formation and cycle behavior of Sn battery anodes. *Journal of Power Sources* **2022,** *535*, 231439.
14. Wang, K.; Joshi, Y.; Chen, H.; Schmitz, G., Quantitative investigation of the cycling behavior and SEI formation of tin through time-resolved microgravimetry. *Journal of Power Sources* **2023,** *569*, 232919.
15. Wang, K.; Joshi, Y.; Kohler, T.; Mead, M.; Schmitz, G., Reversible interfacial Li-oxide formation on germanium and silicon anodes revealed by time-resolved microgravimetry. *Journal of Materials Chemistry A* **2024,** *12* (8), 4610-4622.

Supplementary information for

# Is Platinum a Proton Blocking Catalyst?

*Aparna Saksena*[*a], *Yujun Zhao*[a], *J. Manoj Prabhakar*[a], *Dierk Raabe*[a], *Baptiste Gault*[a,b], *Yug Joshi*[*a]

[a] Max Planck Institute for Sustainable Materials, Max-Planck-Straβe 1, Düsseldorf 40237, Germany

[b] Department of Materials, Royal School of Mines, Imperial College London, Prince Consort Road, London, SW7 2BP, UK

*corresponding author: a.saksena@mpi-susmat.de, y.joshi@mpi-susmat.de

## Atom probe Tomography (APT)

For APT measurements and D analysis of the specimen extracted from Pt$_{DER}$, the probability of molecular ions detected at 2 and 3 Da cannot be neglected. The contribution to these peaks is convoluted as for peak at 2 Da either/both $H_2^+$ molecular ion or/and $D^+$ ion is possible. Similarly peak at 3 Da can have contributions from either/both $H_3^+$ molecular ion and/or $DH^+$ ion. This depends on the instantaneous field conditions, where the charge state ratio (CSR) of the detected Pt can be used as a proxy. To accurately estimate the observed D in the Pt$_{DER}$ specimen that underwent DER at a constant current density of -2.4 mA·cm$^{-2}$ for 12 h, reference Pt specimens are extracted from pristine Pt samples. The field condition is varied by changing the laser pulse energy from 70 pJ to 150 pJ, where the overall mass spectra obtained from 3 such specimens are shown in

Fig. S1. Since these specimens are not influenced by any D interaction, the peak at 2 Da exclusively corresponds to the field evaporation of $H_2^+$ molecular ion.

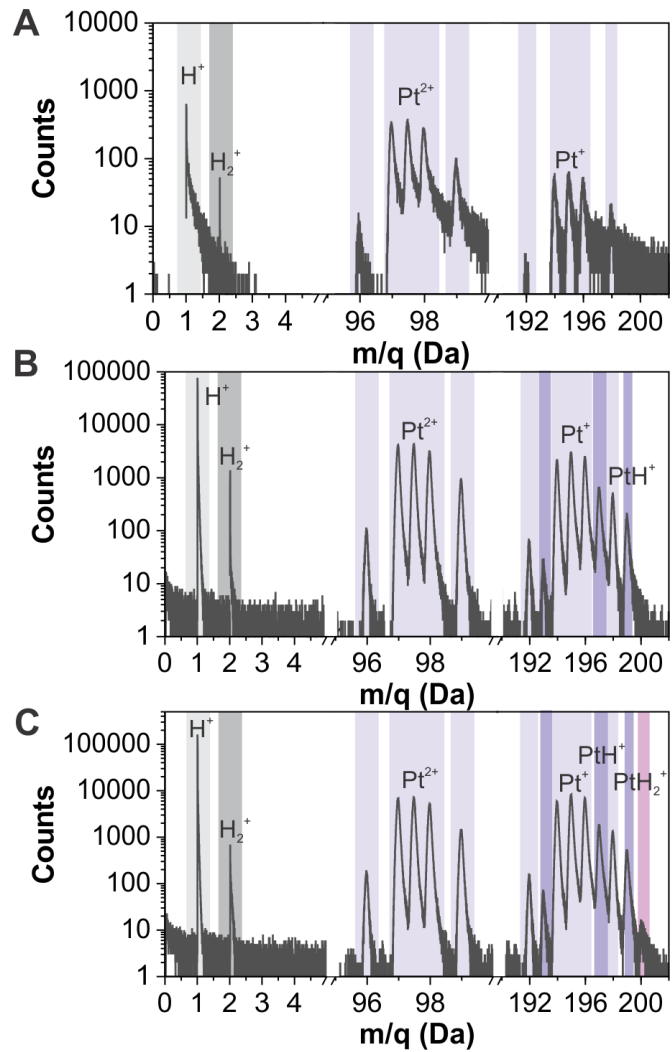

Fig. S1 shows the overall mass spectrum collected from Pt reference specimens, where the field condidtions are varied by changing the applied laser pulse energy.

Based on the reference measurements shown in Fig. S1, log of CSR of $Pt^{1+}$ is plotted as a function of the relative concentration of the $H_2^+$ with respect to the detected $H^+$, shown in Fig. S2A, represented by black open circles. Also, for this quantification only the first isotope of Pt, for both charge states, is considered to avoid any influence of the detected $PtH^+$ ions. The quantitative values of the instantaneous field are shown at various CSRs which are extracted from the calculated Kingham plots[1] for Pt. Similar analysis is performed for the Pt$_{DER}$ specimen, shown in Fig. S2A with red squares. Field conditions similar to Pt$_{DER}$ specimen is successfully reproduced by the reference measurements, where overlap between 26.5 V·nm$^{-1}$ and 29.3 V·nm$^{-1}$ can be clearly observed. It is evident that at similar field conditions, Pt$_{DER}$ specimen shows much higher 2 Da/H$^+$ ratio compared to the reference measurements, clearly indicating contributions of D within this peak. At these field conditions, no peak at 3 Da is observed in the reference measurements which also indicates the sole contribution from DH$^+$ ions in the case of Pt$_{DER}$. These findings allowed us to correct the one-dimensional concentration profile extracted from the Pt$_{DER}$ specimen, where the unprocessed concentration is shown in Fig S2B. The contributions from peaks at 3 Da and 4 Da is shown in Fig S2C along with the expected H$_2^+$ ion contributing to the peak at 2 Da. The expected H$_2^+$ ions are estimated by multiplying the measured H$^+$ ion concentration in the Pt$_{DER}$ specimen by the 2 Da/1 Da ratio obtained from the reference measurements for the corresponding field conditions. Figure S2C also shows the background hits that is estimated by ranging the region in the vicinity of a peak (at 2, 3 and 4 Da) in the mass spectrum, of the same width. The final estimation of D as a function of distance is obtained by adding all of the contributions from peak at 2 Da, 3 Da, twice from 4 Da (D$^{2+}$ ion) while subtracting the expected H$^{2+}$ ion and background associated for all 3 peaks.

---

[1] Kingham, D. R., The post-ionization of field evaporated ions: A theoretical explanation of multiple charge states. *Surface Science* **1982,** *116* (2), 273-301.

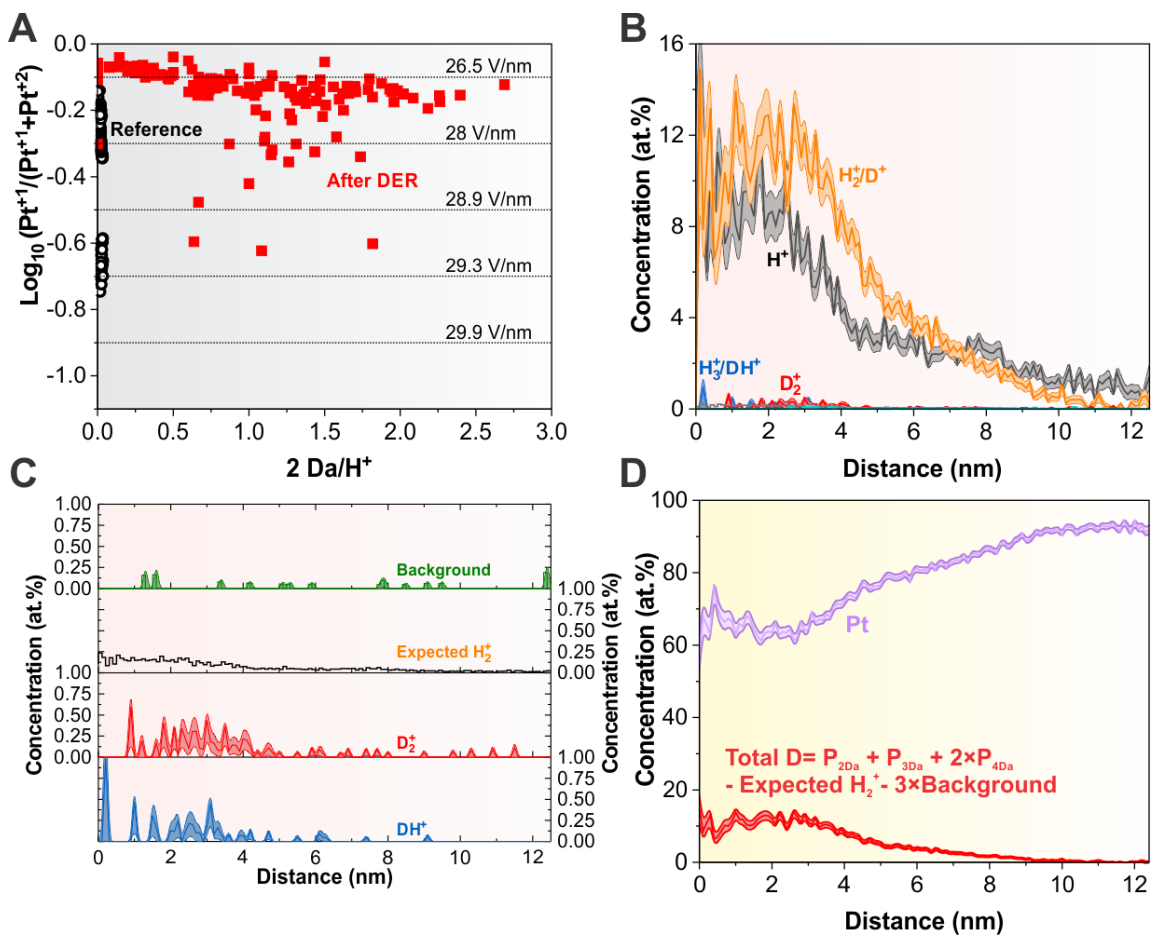

Fig. S2: A) shows the relative concetration of peak at 2 Da with respect to the peat at 1 Da as a function of the field, estimated by the charge state ratio of $Pt^{+1}$ and $Pt^{+2}$. B) shows the unprocessed one dimensional concentration profile of the Pt specimen after DER. C) shows the one dimensional concentration profile of the observed background, expected $H_2^+$ contribution, estimated from the reference data in A), $D_2^+$ and $DH^+$. D) shows the corrected, total D observed in the specimen as a function of distance from the measurement direction.

## Thermal desorption spectroscopy

Fig. S3 shows the thermal desorption spectra where the desorbed hydrogen as a function of temperature is monitored using a mass spectrometer. A heating ramp of 8 °C·min$^{-1}$ is applied. The reference Pt sample is monitored to detect the background and compared to the Pt sample that underwent constant current HER at a current density of -2.4 mA·cm$^{-2}$ for 12 h.

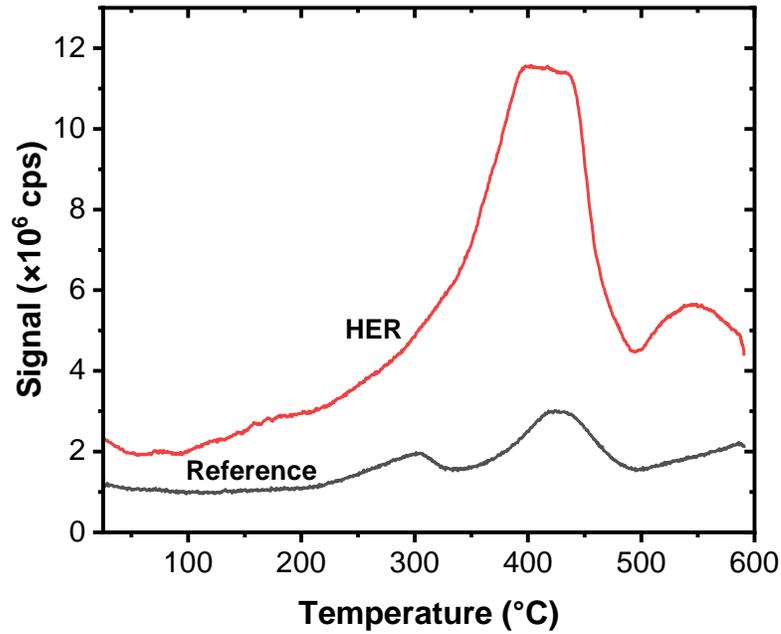

Fig. S3: Thermal desorption spectrum of the reference Pt and after constant current HER

## Error function fitting for D in Pt

The error function used for fitting the composition profile is given in Eq. S1.

$$C(x, D, t) = C_1 + (C_0 - C_1)\mathrm{erf}\left(\frac{x - x_0}{2\sqrt{Dt}}\right) \qquad \text{S1}$$

Where $C$ is the concentration of D in Pt in at. %; $C_1$ is surface concentration; $C_0$ is concentration at infinite distance; $D$ is the diffusion coefficient of D in Pt; t is time; $x$ is the depth of the thin-film. $C_1$ is kept constant to be 50 at% with the assumption that any correction is done automatically via the displacement of the error function by $x_0$ which is a fitting variable; $C_0$ is set to zero as the concentration goes to zero at 10 nm; and t is kept constant at 12 h or 43200 sec.

Calibration curve for the electrochemical Quartz Crystal Microbalance (eQCM)

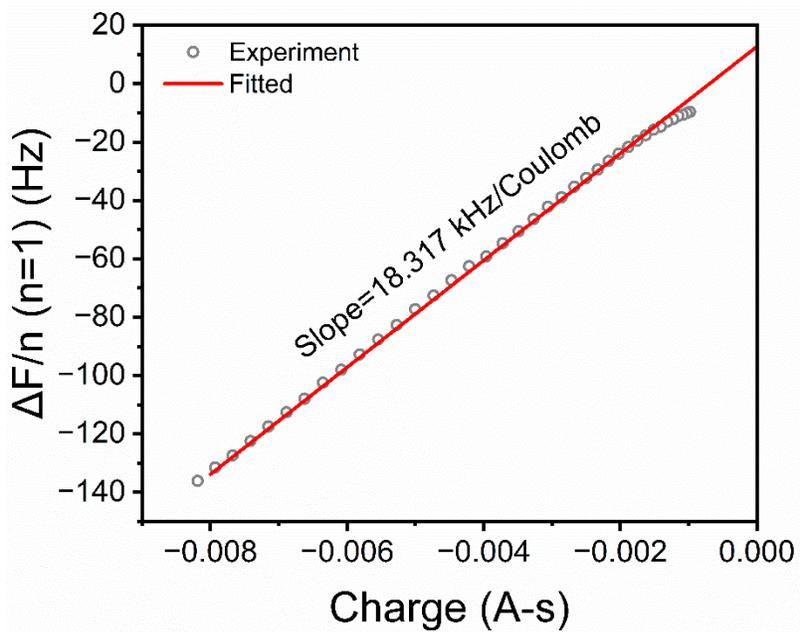

Fig. S4. The calibration curve, it is used to determine the sensitivity factor.